\documentstyle[11pt,psfig]{article}
\begin{document}
\renewcommand{\thepage}{ }
\begin{titlepage}
\title{
\hfill
\vspace{1.5cm}
{\center \bf A single chain analysis of doped quasi one dimensional
spin 1 compounds: paramagnetic versus spin $1/2$ doping}
}
\author{
M. Fabrizio and R. M\'elin\\
{}\\
{International School for Advanced Studies (SISSA-ISAS),}\\
{Via Beirut 2--4, 34014 Trieste, Italy}\\
{and}\\
{Istituto Nazionale per la Fisica della Materia (I.N.F.M.)}\\
{}\\
}
\date{}
\maketitle
\begin{abstract}
\normalsize
We present a numerical study of single chain models of 
doped spin 1 compounds.
We use low energy effective one-dimensional models for both
the cases of paramagnetic and spin-$1/2$ doping. In the case
of paramagnetic doping, the effective model is equivalent
to the bond disordered spin-$1/2$ chain model recently analyzed by
means of real space renormalization group by Hyman and Yang. 
By means of exact diagonalizations in the XX
limit, we confirm the stability of the Haldane phase for weak
disorder. Above a critical amount of disorder, the effective
model flows to the so called random singlet fixed point.
In the case of spin-$1/2$ doping, we argue that the Haldane phase
should be destabilized even for weak disorder. This picture is
not in contradiction with existing experimental data.
We also discuss the possible occurrence of (unobserved)
antiferromagnetically ordered phases.
\end{abstract}
\end{titlepage}

\newpage
\renewcommand{\thepage}{\arabic{page}}
\setcounter{page}{1}
\baselineskip=17pt plus 0.2pt minus 0.1pt
\section{Introduction}
Much effort has been devoted over the last decades to understand
the behavior of quasi one-dimensional spin systems. These 
systems exhibit a rich variety of phases due to the enhanced quantum
fluctuations in reduced dimensionality. 
One of the most famous and studied one dimensional property is the so called
Peierls and spin-Peierls instability. Organic compounds exhibiting such
an instability were synthesized already in the '70s \cite{organic}.
The more recent discovery of the inorganic
CuGeO$_3$ spin-Peierls compound
\cite{CuGeO3,Hase} with a spin-Peierls temperature up to about
15 K renewed the interest to spin-Peierls systems, especially
to the role of various possible doping mechanisms. In fact, 
it is possible to dope these inorganic compounds in a very controlled fashion
with both magnetic and paramagnetic ions
(see for instance \cite{Si-a,Si-b} for Si doping,
\cite{Zn-a,Zn-b,Zn-c} for Zn
doping and \cite{Ni} for Ni doping). It turns out that
an antiferromagnetic (AF) phase is induced even for very
small impurity concentrations, with a maximal N\'eel
temperature of the order of 4 K at few percent doping.
Neutron scattering experiments \cite{Si-b,Zn-b}
have unambiguously identified this phase as an antiferromagnet
with a coexistence of gapless antiferromagnetic
excitations at low energy and spin-Peierls excitations 
at higher energy \cite{Si-b,Zn-c}, and also with
important disorder effects \cite{Zn-b}. In a recent work,
the present authors have shown that insight into the physics
of these compounds can be gained by a single chain model
\cite{Nous1,Nous2}, which does exhibit 
an enhancement upon doping of low energy AF fluctuations, coexisting with 
higher energy spin-Peierls features.
More specifically, by means of the renormalization procedure
proposed in \cite{RG-1},
it was shown in \cite{Nous2} that this effective
model belongs to the Random Singlet (RS) universality class of a random
Heisenberg model \cite{RG-2}. Although within a single chain model 
it is difficult to handle rigorously the interchain
couplings that stabilize the N\'eel phase, nevertheless this approach
does emphasize the conjugated effects of low dimensional quantum 
fluctuations and disorder, which seems to be an important ingredient
in the physics of these compounds.

Another type of low dimensional quantum phase of
spin systems is the Haldane gaped phase \cite{Haldane} of
a spin-1 antiferromagnetic chain, which has been  
observed in quasi one
dimensional compounds \cite{Exp}. Quite interestingly, 
a spin-1 chain and a dimerized
spin-$1/2$ chain belong to the same quantum phase, with a non
vanishing string order parameter \cite{Hida}. However, as we
are going to argue, in spite of this analogy,  the spin-Peierls and 
spin-1 chains behave very differently upon doping.
Moreover, paramagnetic doping of spin-1
chains is very different from spin-$1/2$ doping.
In fact, there are some experimental evidences that
the Haldane phase is robust against paramagnetic doping:
it was found in \cite{Dis-1} that the Haldane gap persists
up to 20$\%$ doping in the NiC$_2$O$_4$DMIZ compound
where Ni was substituted by Zn. On the other hand,
it was shown in \cite{NENP} that doping NENP compounds
with less that 1$\%$ of Cu leads to a Curie behavior in
the susceptibility, with a Curie constant $4.6$ times
larger than the one of the free Cu$^{2+}$ ions. These
experiments thus show a very different behavior of
quasi one-dimensional spin-1 compounds upon doping:
weak paramagnetic doping preserves the Haldane gaped phase
whereas weak spin-$1/2$ doping induces low lying energy states. 
We argue that the behavior of quasi one dimensional spin-1
systems upon both spin-$1/2$ and paramagnetic doping
can be understood on the same
basis as the doped spin-Peierls compounds,
namely, in the framework of a one dimensional
effective model with a rigorous treatment of disorder effects.
We also examine the issue of whether a strong doping
antiferromagnetic phase could appear for paramagnetic doping.

This article is organized as follows.
In Section \ref{ZeroTpara}  we introduce an effective one dimensional
model for paramagnetic doping. The zero temperature phases
of this model were analyzed recently \cite{Hyman,Monthus}.
We check the nature of the zero temperature phases
using exact diagonalizations
of the effective spin-$1/2$ model at zero temperature in the
XX limit. We then argue that, in close analogy to our approach for
spin-Peierls compounds, the zero temperature Haldane phase
of the effective model should be robust against switching
interchain couplings whereas the RS phase should
possibly turn into an antiferromagnetic phase.
Section \ref{ZeroT1/2} is devoted to an analysis of an effective
model for spin-$1/2$ doping. We show that in this case the
Haldane phase is destabilized by an infinitesimal doping.
In section \ref{Exact}, we analyze the question whether
an antiferromagnetic phase might appear for large paramagnetic
doping. We carry out
exact diagonalizations of small clusters, the interchain
coupling being treated in mean field. These diagonalizations
support our prediction, namely the possible
emergence of a strong disordered antiferromagnetic phase.
Finally, section \ref{conclusions} is devoted to some
final remarks.

\section{Effective one dimensional model
for paramagnetic doping}
\label{ZeroTpara}
\subsection{The model}
Based on the Valence Bond (VB) description of a spin-1 chain 
\cite{VB}, Hyman and Yang \cite{Hyman} first argued that bond-disordered
spin-1 chains can be described by the
following low energy effective model: introducing bond disorder
amounts as a first approximation to ``break'' the chain into clusters of 
consecutive strongly coupled spins. As a consequence, effective
low energy spin-$1/2$ degrees of freedom appear at the edges
of these segments \cite{Affleck}. Two neighboring 1/2-spins, one
at the right and the other at the left edges of two
consecutive segments  are then assumed to be weakly AF coupled.

This model should be also appropriate to describe the effects of 
doping by paramagnetic ions. In fact, if a magnetic ion is substituted by a 
non magnetic impurity
(as in the case of Zn doping in NiC$_2$O$_4$DMIZ \cite{Dis-1}),
one expects that two spin-1/2 edge excitations appear
at the right and the left of the impurity, which get coupled by an 
AF exchange $J_2$ weaker than the bulk exchange $J_H$.
In addition, two spin-1/2 moments at the opposite edges of a
segment are also coupled by virtual polarization of the spin-1
background. Assuming that these two edge moments ${\bf S_1}$
and ${\bf S_2}$ are at distance $l$, the effective interaction
mediated by the spin-1 background is
\cite{Affleck}
\begin{equation}
{\cal H}_{1,2}(l) \sim J_H (-1)^{l} e^{-l/\xi_0} {\bf S}_1.{\bf S}_2 
,
\end{equation}
where we have discarded a $1/\sqrt{l}$ prefactor, irrelevant to the
physics we want to discuss.

\subsection{Exact diagonalizations in the XX limit}
\label{exactXX}
Hyman and Yang \cite{Hyman}, and also Monthus
{\it et al} \cite{Monthus},
carried out a RG analysis of this effective model.
These authors showed that,
for weak disorder, the system remains in the gaped Haldane phase.
As the disorder increases above a critical value, 
the system exhibits a second order transition to
the RS phase. The aim of this section
is to check this prediction by means of 
exact diagonalizations in the XX limit. 
Due to the presence of ferromagnetic bonds, the XX limit is not
expected to give a good description of the isotropic XXX model,
contrary to what is found when all bonds are AF.
However, like the XXX model analyzed in Ref. \cite{Hyman}, also the
XX version may have in principle two different phases for weak and 
strong doping. Therefore, if one is only interested in confirming 
the existence of this phase transition, the XX model should be sufficient. 

In the XX limit, this model can be maped onto
non interacting spinless fermions via a
Jordan-Wigner transformation \cite{Lieb}:
\begin{equation}
{\cal H} = \frac{1}{2} \sum_i J_i \left(
c_{i+1}^{+} c_i + c_i^{+} c_{i+1} \right)
,
\label{effective}
\end{equation}
where the label $i$ denotes the sites carrying spin $1/2$ moments
and $J_i$ is calculated according to the aforementioned rules.
We restrict ourselves to an even number of spin $1/2$
magnetic moments, so that the fermions are periodic
and the Fermi sea is half-filled.
Moreover, the impurities are distributed according to
a Poisson law with an average impurity to impurity
distance $\langle l \rangle$.

Increasing the strength of disorder can be done
either by decreasing 
$J_2$ or by reducing the average impurity to impurity
distance $\langle l \rangle$. We have found that in both cases 
increasing the strength of disorder
leads to an increase of the correlation length and
finally to power law correlations. 
We now present our calculations.

\begin{figure}
\centerline{\psfig{file=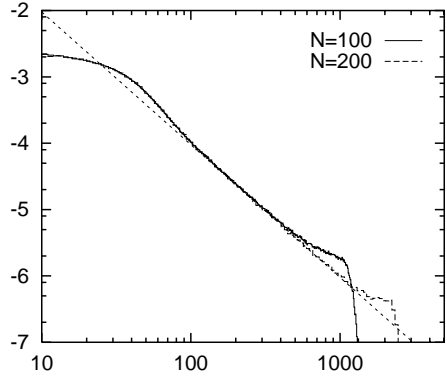,width=5cm}}
\caption{Logarithm of the correlation
between the $z$ component of the
spins in the strong disorder phase (paramagnetic doping),
as a function of the spin-spin distance.
The AF exchange between spins 1 is $J_H=1$,
the nearest neighbor AF exchange is $J_2=0.1$,
the correlation length in the Haldane phase is $\xi_0=20$
lattice spacings, the average distance between two impurities
is $\langle l \rangle = 50$ lattice spacings. For $N=100$ ($200$)
impurities, we performed the average over $32000$ ($2000$) disorder
realizations. The straight line is a fit to a $1/R^{2}$ behavior.
}
\label{Fig1}
\end{figure}

In order to calculate the $S_0^{z} S_R^{z}$ correlations,
we first diagonalize the tight binding Hamiltonian
\begin{equation}
{\cal H}^{(TB)} = \frac{1}{2} \sum_i J_i \left( |i+1 \rangle 
\langle i | + |i \rangle \langle i+1 | \right)
\label{Htb}
.
\end{equation}
If $\Psi_{\alpha} = \sum_i \Psi^{\alpha}_i |i\rangle$
denotes the eigenstates of (\ref{Htb}), the average correlations
are $\langle S_i^{z} S_j^{z} \rangle = A_{i,j} - B_i/2
- B_j/2$, with
\begin{eqnarray}
A_{i,j} &=& \sum_{\alpha,\beta \in FS} \left(
\Psi^{\alpha}_i \Psi^{\beta}_j \right)^{2} 
+ \sum_{\alpha \in FS} \sum_{\beta \not{\in}
FS} \Psi_i^{\alpha} \Psi_i^{\beta}
\Psi_j^{\alpha} \Psi_j^{\beta}\\
B_{i} &=& \sum_{\alpha \in FS} \left(
\Psi_i^{\alpha} \right)^{2}
.
\end{eqnarray}
The correlations $\overline{\langle S_0^{z} S_R^{z} \rangle}$
are shown on Fig. \ref{Fig1} in the strong disorder regime.
This correlation function follows quite nicely a 
$1/R^{2}$ decay, as in
the random singlet phase \cite{RG-2}.

\begin{figure}
\centerline{\psfig{file=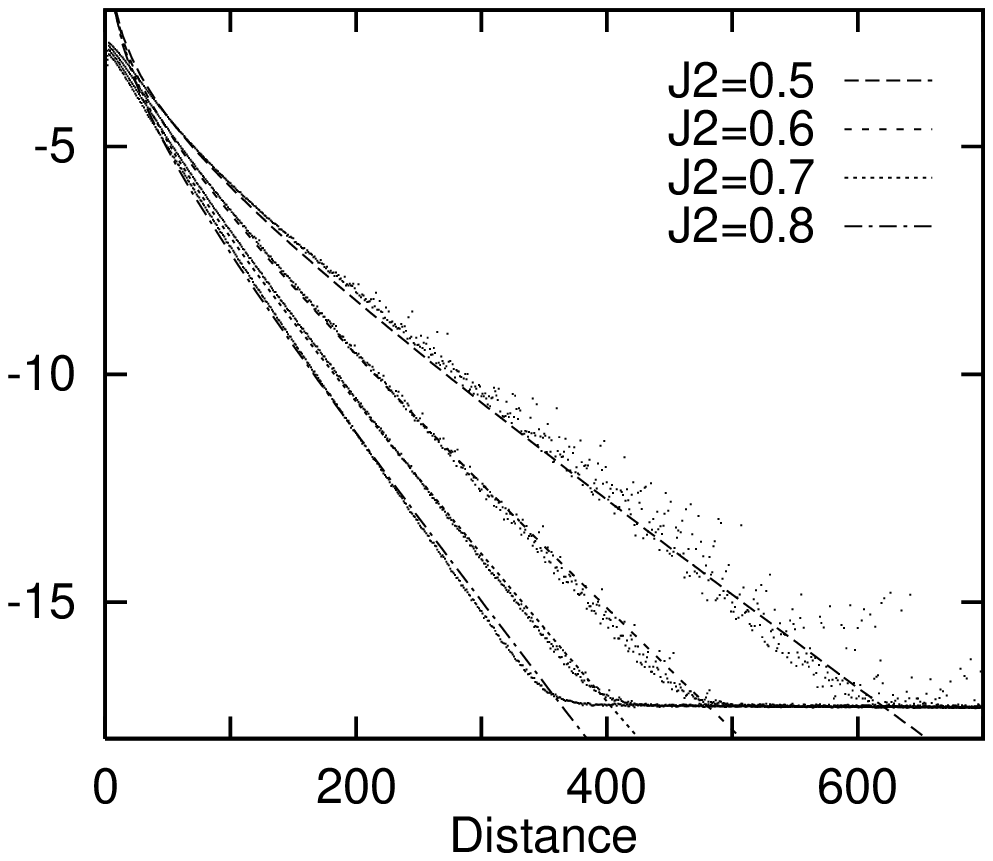,width=4.5cm}}
\caption{Logarithm of the correlation of the $z$
component of the spins
in the small disorder phase (paramagnetic doping).
This is a semi-log plot.
The AF exchange between spins 1 is $J_H=1$;
the nearest neighbor AF exchange is $J_2=0.5,0.6,
0.7,0.8$;
the correlation length in the Haldane phase is $\xi_0=20$
lattice spacings; the average distance between two impurities
is $\langle l \rangle = 50$ lattice spacings. The correlations
have been calculated for $N=100$ impurities and averaged respectively
over $50000$, $10000$, $6000$ and $1300$ realizations of the disorder.
The solid lines are a fit to
the form $\exp{(-R/\xi)}/R^{2}$, with respectively
$\xi=23 \pm 0.5$, $17.5 \pm 0.5$, $14.5 \pm 0.5$, $13 \pm 0.5$.
The saturation for correlations
smaller than $\simeq 10^{-19}$ is due to round-off errors during
the diagonalizations.
}
\label{Fig2}
\end{figure}

We now keep a constant dilution of impurities but increase
the nearest neighbor interaction $J_2$ across the impurities.
This should drive the systems into the Haldane phase, with exponentially
decaying correlation functions. This is clearly seen for the $S^{z}_0
S^{z}_R$ correlation plotted in Fig. \ref{Fig2} 
in a semi-log plot for different values of $J_2$.
Notice on Fig.\ref{Fig2} an increase of the scattering of the data
as the system approaches the zero temperature critical point,
in spite of an increase
of the number of disorder realizations. This is a clear signature
of critical fluctuations and thus consistent with the prediction
that this zero temperature transition is a second order
transition \cite{Hyman,Monthus}.

\begin{figure}
\centerline{\psfig{file=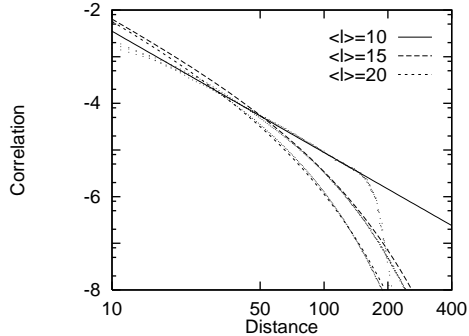,width=4.5cm}}
\caption{Logarithm of the correlation of the $z$ component of the spins
in the regime $\langle l \rangle < \xi_0$ (paramagnetic doping).
This is a log-log plot.
The AF exchange between spins 1 is $J_H=1$,
the nearest neighbor AF exchange is $J_2=0.7$
the correlation length in the Haldane phase is $\xi_0=20$
lattice spacings, the average distance between two impurities
is $\langle l \rangle = 10,15,20$ lattice spacings. The correlations
have been calculated for $N=100$ impurities and averaged respectively
over $3400$, $10000$, $25000$ realizations of the disorder.
The solid lines are a fit to
the power law $1/R^{2.4}$ in the case
$\langle l \rangle=10$; to the form
$\exp{(-R/\xi)}/R^{2.4}$ with $\xi=45$ lattice spacing
for $\langle l \rangle = 15$; and $\xi = 30$ lattice spacings for
$\langle l \rangle = 20$.
}
\label{Fig3}
\end{figure}

We now consider the case $J_2=0.7$ on Fig. \ref{Fig2} and
drive the system towards the infinite correlation length
phase by decreasing the average impurity to impurity distance
$\langle l \rangle$.
The resulting correlations are plotted on Fig. \ref{Fig3}.
We clearly see that decreasing the average distance between impurities
drives the system towards a phase with critical correlations.

Although the effective model is valid only if $\langle l \rangle > \xi_0$,
we have decided to investigate also the situation
$\langle l \rangle < \xi_0$ just as a theoretical model. 
In this case we were
not able to fit the correlation to the RS $1/R^{2}$ form. Instead,
we found in the case of Fig. \ref{Fig3} a power law decay
different from the RS one. Indeed, we find
\begin{equation}
\overline{\langle S_0^{z} S_R^{z} \rangle} \sim
\frac{1}{R^{2.4}}
.
\end{equation}
This may be a finite size effect:
we know that the renormalization group procedure is asymptotically
correct since the disorder distribution becomes extremely large
as the decimation is carried out. In the case
$\langle l \rangle < \xi_0$, it
is possible that, given the small sizes that we use, the
distribution of bonds is not large enough, so that the
correlation exponent has still not converged
the one of the RS. Also, we notice that
the bond distribution is
\begin{equation}
P(|J|) = \frac{1}{2 J_H}\frac{\xi_0}{\langle l \rangle}
\left(\frac{ |J|}{J_H} \right)^{-1+\frac{\xi_0}{\langle l \rangle}}
\theta \left( J_H-|J| \right) + \frac{1}{2} \delta(J-J_2)
.
\end{equation}
In the case $\xi_0<\langle l \rangle$, the smaller the exchange
the larger the probability, which explains that in this regime the
numerics works so well (see Fig.\ref{Fig1}). In the opposite
regime ($\xi_0 > \langle l \rangle$), the smaller the
exchange the smaller the probability. In this case,
we still expect the RS fixed point.
However, it is not so obvious
to exhibit it from a numerical point of view.

\section{Effective one dimensional model for spin $1/2$ doping}
\label{ZeroT1/2}
\subsection{Effective one dimensional model}
The effect of spin $1/2$ doping in quasi one dimension spin 1
NENP compounds was analyzed in \cite{NENP} where susceptibility
measurements were carried out, and also in \cite{Affleck}
by means of ESR measurements. These last experiments showed
unambiguously that spin $1/2$ doping amounts to release
two spin $1/2$ effective moments on the right and the left
of the impurity, that are weakly AF coupled
to the spin $1/2$ impurity. This picture is not in contradiction
with the susceptibility measurements \cite{NENP}, where
the temperature variations of the susceptibility were found
to have Curie behavior. 
The effective model thus consists of disordered units of
three consecutive spin $1/2$ spins. The spin $1/2$ moments
experience a weak AF interaction $J'$
and two consecutive
units are coupled via the ferromagnetic or AF
interaction (\ref{effective}).

It is immediately clear that
the RG description of this effective model is very different
from the one in \cite{Hyman}. The reason is that, in the weak
disorder case, one first would replace all the 3 spin $1/2$
units by a spin $1/2$. Therefore the effective model would
be again a bond disordered $S=1/2$ Heisenberg AF,
but the bonds
being either ferromagnetic or AF. This model
has been investigated in \cite{Leeetal} quite in detail
and shown to be unstable against a small disorder.
In particular, the ground state is expected to have
power law correlations. We check in section \ref{numerics}
that our effective model indeed has power law correlations in its
ground state. However, the physics is expected to be very
different from the RS: the low temperature susceptibility
has a Curie dependence whereas in the RS the low temperature
susceptibility
diverges slower  than a Curie behavior ($\chi(T) \sim 1/(T\ln^{2}{T})$)
\cite{RG-2}. Moreover, the low temperature physics is dominated
by large effective spins. Coming back to the physics of
spin $1/2$ doped quasi one dimensional spin 1 systems, we
do not know whether, given this effective description,
it would be possible to drive the system to a N\'eel phase
by switching interchain couplings.
\subsection{Exact diagonalizations in the XX limit}
\label{numerics}
\begin{figure}
\centerline{\psfig{file=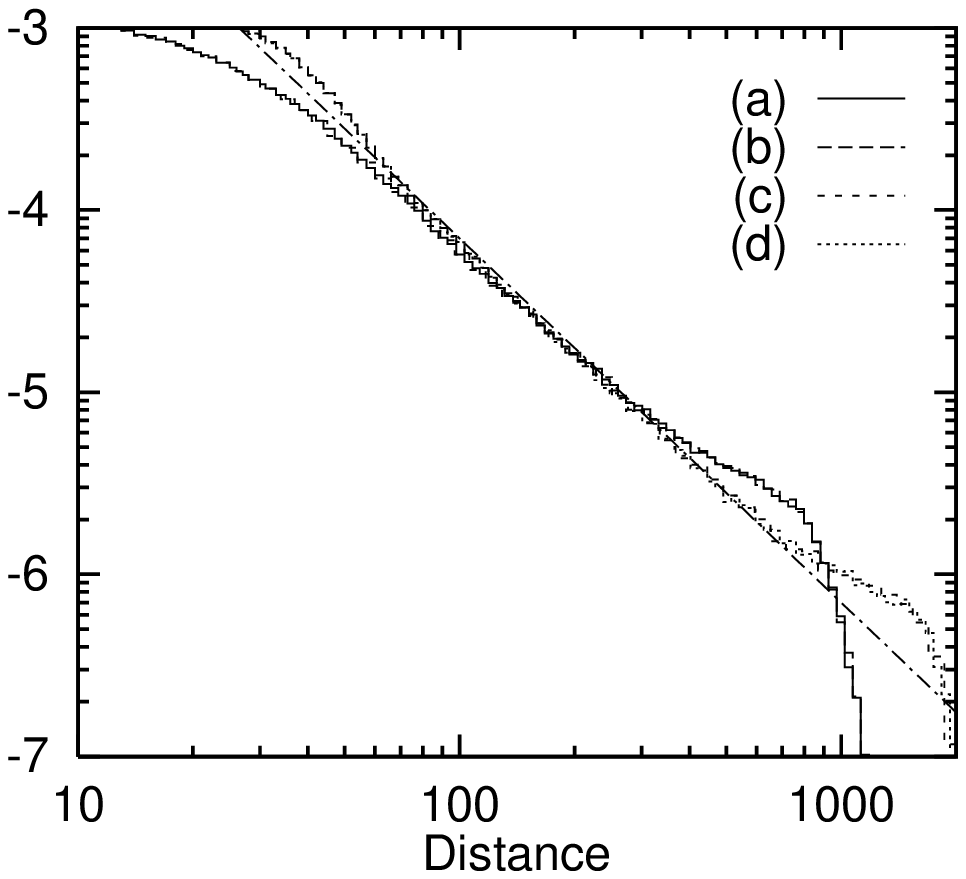,width=5cm}}
\caption{Logarithm of the correlations of the $z$ component of the spins
for the effective model of spin $1/2$ doping. The data
have been fited to a $1/R^{2}$ dependence.
we have chosen: $J_H=1$, $\xi_0=20$,
$\langle l \rangle=50$ and
(a): $J'=0.8$, $102$ spin $1/2$ moments
(b): $J'=0.1$, $102$ spin $1/2$ moments.
(c): $J'=0.8$, $204$ spin $1/2$ moments.
(d): $J'=0.1$, $204$ spin $1/2$ moments.
}
\label{Fig4}
\end{figure}

We repeat the numerical calculations of section
\ref{exactXX} in the case of the effective model for
spin $1/2$ doping. As in the case of paramagnetic doping,
we only consider the spin anisotropic XX model.
The $\overline{\langle S_0^{z} S_R^{z} \rangle}$ correlations
are plotted on Fig. \ref{Fig4} for weak and strong disorder
and exhibit a $1/R^{2}$ power law decay, like for the RS phase. 
This numerical calculation thus supports the previous argument
that the effective one dimensional model suited for spin $1/2$ doping
would flow to a massless fixed point as soon as disorder
is introduced.

\section{Exact diagonalizations of small clusters with paramagnetic doping}
\label{Exact}
\begin{figure}
\centerline{\psfig{file=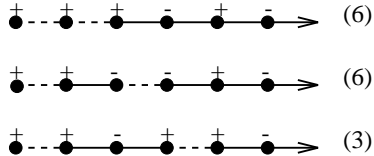,width=2cm}}
\caption{The three inequivalent disorder realizations,
represented in the six sites squeezed chain. Arrows denote periodic
boundary conditions. A dashed line denotes a weak bond.
The sign of the staggered field is indicated by a $\pm$ sign.
The degeneracy is indicated in brackets.
}
\label{Fig5}
\end{figure}

\begin{figure}
\centerline{\psfig{file=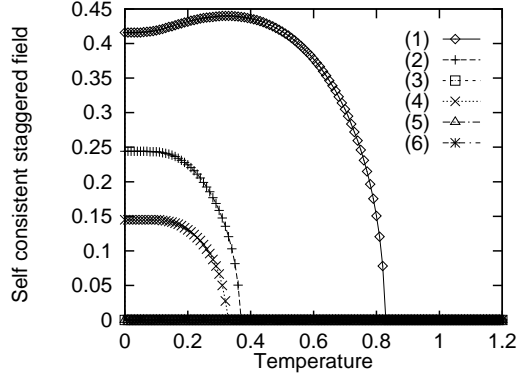,width=5cm}}
\caption{Self consistent staggered field as a function
of temperature calculated for $J_H=1$, $J_2=0.5$
$J_{\perp}=.8$ ((1),(2) and (3))
and $J_{\perp}=0.4$ ((4),(5),(6)). (1) and (4), (2) and (5),
(3) and (6)  correspond to the disorder realizations No 1, 2, 3
on Fig. \protect\ref{Fig5}.
}
\label{Fig6}
\end{figure}

We now turn to exact diagonalizations of small clusters at finite
temperatures, the interchain coupling being treated in mean field.
Notice that this problem is quite difficult and we cannot go
to quite large sizes because (i) one must calculate {\it all}
the eigenvalues and eigenvectors since we are interested
in the finite temperature behavior, (ii) one must calculate
the staggered field in mean field, which implies solving for
self consistent equations, (iii) one must average
over disorder. This kind of calculation was already carried
out in the context of disordered spin-Peierls systems \cite{Nous2},
with results not in contradiction with experiments.
We consider an eight site chain with two paramagnetic impurities,
and thus six sites carrying a spin 1 in the squeezed chain.
There are three inequivalent
ways of putting the impurities on the chain as shown on Fig.
\ref{Fig5}.
We apply a finite
staggered field and calculate the staggered magnetization response
for different magnetic fields and temperatures. In practice,
the temperature and the staggered field are varied between
$0.01$ and $2$ with an increment of $0.01$ and finally the
self consistent staggered field $h_s^{*}$ is calculated
as a function of temperature
by imposing the condition $h_s^{*}(T) = J_{\perp}
m_s(T,h^{*}_s(T))$, where $J_{\perp}$ is the
strength of the effective interchain coupling.
The self consistent staggered field is plotted on Fig.
\ref{Fig6} for two values of the effective
interchain coupling $J_{\perp}$.
As in the case of the spin-Peierls system, we observe of Fig.
\ref{Fig6} important fluctuations of the mean field N\'eel
transition temperature. At this point, we conclude that
a transition to an AF ordered state should be possible
in the strong disorder regime.

\begin{figure}
\centerline{\psfig{file=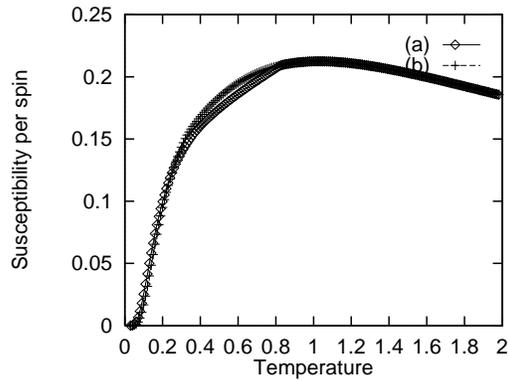,width=5cm}}
\caption{Variations of the uniform susceptibility per spin
as a function of temperature for $J_H=1$, $J_2=0.5$ and
(a): $J_{\perp}=0.8$ (b): $J_{\perp}=0.4$.
}
\label{Fig7}
\end{figure}

In order to have an idea of the variations of a macroscopic
observable, we plotted on Fig.\ref{Fig7} the variations of
the spin susceptibility as a function of temperature for
two values of the interchain coupling.
We observe that unlike the case of the spin-Peierls system
\cite{Nous2}, the emergence of AF as the temperature
is lowered does not induce a maximum of the susceptibility,
but a jump in the slope of the susceptibility versus temperature.
 
Moreover, as a comparison between the two different doping mechanisms
in a spin-1 chain, we note that
eliminating the non magnetic sites going from the
original chain to the squeezed chain has more drastic consequences
for the paramagnetic doping than for the spin $1/2$ doping. 
As far as paramagnetic doping is concerned, applying a
staggered field in the original chain amounts to apply a staggered
field in the {\it same} direction on both the spin $1/2$ moments
released by the impurity (on the right and the left of the impurity)
and thus to systematically frustrate the AF coupling 
 $J_2$ across the impurity. Notice that in the case
of spin $1/2$ doping, this competition does not arise since the
spin $1/2$ impurity is coupled by AF bonds to
the localized effective spin $1/2$ moments. Hence, in the case of
paramagnetic doping, this effect should play against
the establishment of an AF phase.

\section{Conclusions}
\label{conclusions}
We have thus carried out a detailed investigation
of the effect of doping spin 1 chains by paramagnetic
and spin $1/2$ impurities.
Our approach takes advantage of the quasi one dimensionality
of these systems. We have shown how these systems can be maped
onto bond disordered spin 1/2 chains, the zero temperature fixed
points of which were recently analyzed (\cite{Hyman,Monthus} for
the effective model appropriate to paramagnetic doping, and
\cite{Leeetal} for spin $1/2$ doping).
We have pointed out that paramagnetic and spin $1/2$ doping
are different physical situations since the effective one
dimensional models have a very different physics. Weak paramagnetic
doping preserves the Haldane phase \cite{Dis-1}. Above a
critical disorder strength, the effective one dimensional model
flow to the RS phase. 
On the basis of our previous experience of spin-Peierls
systems \cite{Nous1,Nous2},
we argue that the RS phase, which was found to be the 
strong disorder zero temperature fixed point,
would translate in these paramagnetically doped
spin 1 chains
into a possible AF phase above a critical disorder.
Quasi one dimensional spin 1 compounds
were already found to be robust against
paramagnetic doping \cite{Dis-1}, but, to the best of the authors' knowledge,
no AF ordered phase was found up to now.
We have shown by means of exact diagonalizations of small clusters
that a strong disorder AF phase is possible.
In fact, the disorder favors the
enhancement of magnetic fluctuations (as in the case of
spin-Peierls compounds).
Moreover, we have pointed out that in the paramagnetically
doped spin 1 compounds,
the interchain coupling always frustrates the
AF coupling across the impurity. This mechanism does not exist 
in the spin $1/2$ doped compounds, which in addition flow to a gapless
phase for arbitrary weak disorder.

Although the spin 1/2 doping seems at first sight more promising, 
it is not clear at this stage whether it should or not display an
AF phase when interchain coupling is allowed,
since the physics of the effective one dimensional model
is dominated by clusters of spins coupled ferromagnetically,
or, equivalently, by large spins.

On the other hand, a very difficult problem in the
possible observation of an AF phase in the paramagnetic
doping case is that the microscopic parameters should be tuned
in such way that AF occurs for not too large
impurity concentrations, when the compounds become unstable.
This means that the AF coupling across the impurity
should be as small as possible. Therefore,
the issue of the appearance of AF upon doping spin-1 quasi
one dimensional compounds requires still further
experimental investigations. 

Finally, even in the absence of AF ordering, neutron scattering
experiments should reveal the presence of low-lying
magnetic excitations along the chain directions,
in both the spin-1 and spin-$1/2$ doping situations. The zero temperature 
fluctuations should be critical for a small spin-$1/2$ doping.
For spin-1 doping, the correlation length should remain finite
below a critical amount of disorder, but the correlation length
should increase with doping. To our knowledge, such measurements
have not been carried out up to now in doped NENP or DMIZ
compounds, even though neutron scattering experiments were performed
in the doped quasi one dimensional transition oxyde
Y$_2$CaBaNi$_{1-x}$Zn$_x$O$_5$ \cite{Bell}.
However, this compound is not properly
described by the type of localized spin models that we have analyzed
here. Nonetheless, the results in \cite{Bell} are interesting
in view of our conclusions since the appearance of
low lying AF fluctuations upon doping is reported.

\underline{Acknowledgements:} The authors acknowledge useful
discussions with J.P. Boucher and L.P. Regnault. The numerical
calculations presented here were carried out on the computers
of the theory group of Centre de Recherches sur les Tr\`es Basses
Temp\'eratures, Grenoble. This work has been partly supported
by EEC under contract No ERB CHR XCT 940438, and by INFM,
Project HTCS.


\end{document}